\documentclass[prb,showpacs,twocolumn,nobibnotes,epsf,sort&compress,super]{revtex4}

\usepackage{graphicx}
\usepackage{dcolumn}
\usepackage{bm}
\usepackage{SIunits}

\begin{document}
\title{Enhanced Ferromagnetic Ordering in GdBaCo$_{2}$O$_{5.5+\delta}$ Films on SrTiO3 (001) Substrate }
\author{G. Y. Wang, X. Y. Zhou, R. K. Zheng, Y. M. Hu, H. L. W. Chan}
\author{Y. Wang}
\altaffiliation{Corresponding author}
\email{apywang@inet.polyu.edu.hk} \affiliation{Department of
Applied Physics and Materials Research Center, the Hong Kong
Polytechnic University, Hong Kong SAR, China\\}
\date{\today}

\begin{abstract}
The authors investigated the structure and properties of
GdBaCo$_{2}$O$_{5.5+\delta}$ thin films epitaxially grown on
SrTiO$_{3}$ (001) single crystal substrates. The thin films were
found to have a notable remnant magnetization above room
temperature, which is much higher than that observed in bulk
material. Transmission electron microscopy and x-ray diffraction
patterns reveal that phase separation occurs in these films, and
the phase responsible for the enhanced ferromagnetic order is
$a$-oriented. The enhanced ferromagnetic order is attributed to
the enhanced orbital order of Co$^{3+}$ in CoO$_{5}$ pyramids, and
the disappearance of ferromagnetic to antiferromagnetic transition
is explained the stabilization of higher spin state of Co$^{3+}$
in CoO$_{6}$ octahedra.

\end{abstract}


\vskip 300 pt

\maketitle

\section{INTRODUCTION}

Perovskite-related cobalt oxides, e.g.
$R$$_{1-x}$$A$$_{x}$CoO$_{3}$ ($R$ = rare earth metals, $A$ = Ca,
Sr, and Ba), have recently attracted much attention due to their
interesting physical properties such as colossal
magnetoresistance\cite{Briceno}, large thermoelectric
efficiency\cite{Terasaki}, and spin-state transition\cite{Senaris,
Tsubouchi, Fujita, Fita}. In these oxides, CoO$_{6}$ octahedron is
a fundamental structure in which the Co ions may have different
valences (Co$^{2+}$, Co$^{3+}$ and Co$^{4+}$) and spin states (for
Co$^{3+}$: high spin state, HS, $t$$_{2g}^{4}$$e$$_{g}^{2}$,
$S$=2; intermediate spin state, IS, $t$$_{2g}^{5}$$e$$_{g}^{1}$,
$S$=1; low spin state, LS, $t$$_{2g}^{6}$$e$$_{g}^{0}$, $S$=0).
Since the Hund coupling energy $J$$_{H}$ is comparable to the
crystal-field splitting energy $\Delta$$E$ in some doping levels,
the spin state of Co ions can be converted by stimulation of some
external variables such as temperature\cite{Senaris, Tsubouchi}
and pressure\cite{Fita}, which modifies the value of
$J$$_{H}$/$\Delta$$E$.

Very recently, great interests have been paid to layered cobalt
oxides $R$BaCo$_{2}$O$_{5.5+\delta}$,\cite{Maignan, Roy,
Pomjakushina, Motin, Khalyavin, Troyanchuk, Respaud, Taskin,
Frontera, Wu} where $R^{3+}$ and Ba$^{2+}$ ions locate in
alternating planes along $c$-axis due to their large mismatch in
ion radii. In $R$BaCo$_{2}$O$_{5.5+\delta}$, the oxygen content
can be significantly modified by different annealing process,
which tunes the nominal valence of cobalt from 2.5+ to 3.5+. The
oxygen vacancies always appear in rare-earth planes\cite{Frontera,
Maignan}, and turns two neighboring CoO$_{6}$ octahedra to
pyramids. Among these layered cobalt oxides,
GdBaCo$_{2}$O$_{5.5+\delta}$ (GBCO) has been well
studied\cite{Troyanchuk, Respaud, Taskin, Frontera, Wu} and
multiple phase transition has been observed. In none doping
GdBaCo$_{2}$O$_{5.5}$, half of Co$^{3+}$ ions are in octahedra and
the others are in pyramids. The oxygen vacancies were ordered in a
row running along $a$-axis. A spin-state transition is observed at
360 K, associated with a metal-insulator transition, where the
spin state of Co$^{3+}$ ions in octahedra change from HS to LS
with decreasing temperature while those in pyramid remains
IS.\cite{Taskin} Taskin $et$ $al$. found that the IS Co$^{3+}$
ions in pyramids formed ferromagnetic (FM) two-leg ladders running
along $a$-axis due to orbital ordering, and the spins exhibit a
strong Ising-like anisotropy that all the spins point to
$a$-axis\cite{Taskin}. The ground state of GBCO is
antiferromagnetic (AFM) and changes to FM state at 260 K due to
the appearance of Co$^{2+}$/Co$^{4+}$ pairs or the LS to IS/HS
transition of Co$^{3+}$ in some of the octahedra. At 300 K the
compound undergoes a phase transition from FM state to
paramagnetic state, due to the destroyed FM ladders by thermal
fluctuation\cite{Taskin}.

Although many studies have been performed on bulk material, few
works\cite{Kasper,Yuan} has been done on the substrate-induced
strain effects for $R$BaCo$_{2}$O$_{5.5+\delta}$ thin films. In
this report, we fabricated epitaxial GBCO films on SrTiO$_{3}$
(001) substrates and found a strain-enhanced FM order at high
temperature (430 K), which is seldom observed in perovskite cobalt
oxides.

\section{EXPERIMENT DETAILS}
A ceramic target of GBCO was prepared by the solid-state reaction
method in air with stoichiometric Gd$_{2}$O$_{3}$, BaCO$_{3}$ and
Co$_{3}$O$_{4}$. X-ray diffraction (XRD) measurement shows that
the GBCO ceramic target is single phase. GBCO thin films with
thickness of 40 nm and 100 nm were deposited on SrTiO$_{3}$ (001)
single crystal substrates by pulsed laser ablation technique.
During deposition, the substrate temperature was kept at 700
$\celsius$ and the oxygen pressure was kept at 27 Pa. After
deposition, the films were cooled to room temperature in 1000 Pa
O$_{2}$. As-grown 40 nm film (labeled A1) were post-annealed in
floating oxygen at 400 $\celsius$ for three hours (labeled A2),
and then in air at 900 $\celsius$ for 1 hour (labeled A3). The
structure of the thin films was characterized by a Discover D8
x-ray diffractometer equipped with Cu K$\alpha$ radiation and a
Joel 2010F Transmission electron microscopy (TEM). Vibrating
sample magnetometer (VSM) and superconducting quantum interference
device (SQUID) magnetometer were employed in the magnetic
characterizations.

\section{RESULT AND DISCUSSION}

\begin{figure}[ht]
\centering
\includegraphics[width=8cm]{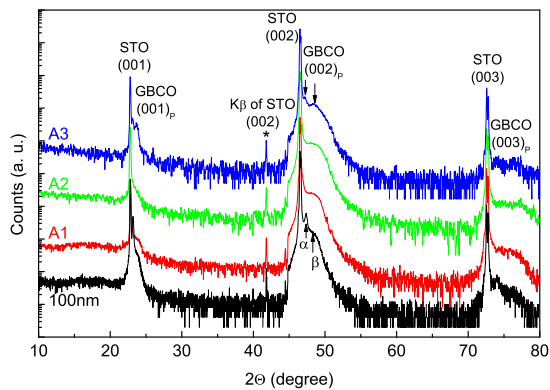}
\caption{X-ray diffraction patterns of the films with different
thickness and heat treatments. The arrows indicate two different
peaks from film.} \label{fig1}
\end{figure}

The GBCO ceramic was found to have an orthorhombic structure with
lattice parameters $a$ = 3.88 \AA, $b$/2 = 3.915 \AA, and $c$/2 =
3.77 \AA, consistent with that reported in
literature\cite{Taskin}. Fig. \ref{fig1} shows the XRD patterns of
thin films with different thickness and heat treatments. For 100
nm film, a peak with strong intensity is found at 2$\Theta$ $\sim$
47.3$^{\circ}$, corresponding to the out-of-plane lattice
parameter (with perovskite structure) 3.85 \AA \mbox{ }of film. A
hump with weaker intensity can be seen at 2$\Theta$ $\sim$ 48.3
$^{\circ}$, indicating smaller out-of-plane lattice parameter of
3.77 \AA. This suggests the possible phase separation in this
film. It should be pointed out that no peak is observed at
2$\Theta$ $\sim$ 12$^{\circ}$ (corresponding to $d$ $\sim$
7.8\AA), which may be caused by the weaker intensity of this peak.
In TbBaCo$_{2}$O$_{5.5+\delta}$ films this peak also disappeared
but appeared after slowly cooling down from high
temperature\cite{Kasper}, indicating the sensitivity of this peak
to the heat treatment in films. For 40 nm films, it can be seen
that all the films show hump at about 48.3$^{\circ}$ , educing the
out-of plane lattice parameter of 3.77 \AA. The hump shows only
small shift among A1, A2 and A3. Comparing with 100 nm film, the
peak at 47.3$^{\circ}$ disappears in A1 and A2, but presents in
A3. This indicates that phase separation is also present in
high-temperature treated A3, which will be confirmed by
high-resolution TEM (HRTEM) images in the last figure. From now on
we defined phase $\alpha$ to the peak with longer out-of-plane
axis (smaller 2$\theta$) and phase $\beta$ to the hump with
shorter out-of-plane axis (larger 2$\theta$).

\begin{figure}[t]
\centering
\includegraphics[width=8cm]{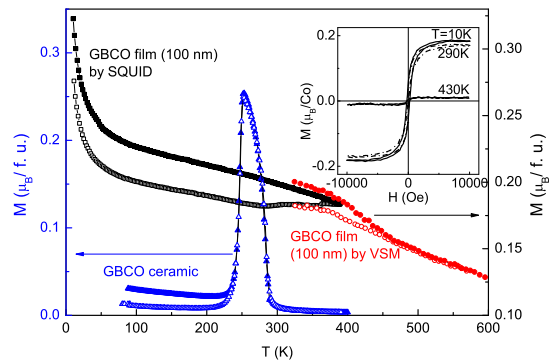}
\caption{Temperature dependence of magnetization for the GBCO
ceramic sample and 100 nm film at H = 1000 Oe. Inset shows the
ferromagnetic component of magnetization for 100 nm film at 10 K,
290 K, and 430 K (the diamagnetic and paramagnetic signals are
subtracted).} \label{fig2}
\end{figure}

Figure \ref{fig2} shows the temperature dependence of
magnetization for the GBCO thin film (100 nm) and ceramic target.
In the ceramic, a FM order emerges at 288 K with decreasing
temperature, but soon changes to AFM order at 255 K. This is a
typical characteristic of GBCO with $\delta$ close to
0.\cite{Taskin} The magnetization of thin film (squares for the
data taken by SQUID and circles for those taken by VSM), however,
is quite different. In film, the FM-AFM transition disappeared,
which could be induced by oxygen deficiency \cite{Taskin} or
substrate-induced strain effect. We note that a remnant
magnetization can be seen even at 430 K, which is higher than that
reported in GBCO ceramic and single-crystal\cite{Troyanchuk,
Respaud, Taskin, Frontera} with any oxygen content. This is an
unusual behavior in perovskite-related cobalt oxides of which the
Curie temperature is usually not higher than 300 K\cite{Frontera,
Maignan, Roy, Pomjakushina, Motin, Khalyavin, Troyanchuk, Respaud,
Taskin}. The inset of Fig. \ref{fig2} shows the field dependence
of magnetization of the thin film at 10, 290 and 430 K,
respectively. The M(H) curves confirm that the FM order emerges at
about 430 K. It should be pointed out that all paramagnetic and
diamagnetic moments have been subtracted and only the FM moment is
presented in the M(H) curves. From this figure, we can see that
the FM order in film is not only expanded to low tempereature, but
also to high temperature comparing with ceramic sample.

\begin{figure}[ht]
\centering
\includegraphics[width=8cm]{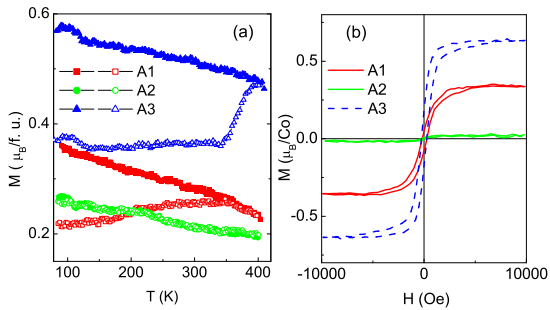}
\caption{(a) Temperature dependence of magnetization at H = 1000
Oe for A1, A2 and A3. Open symbols for zero-field-cooled data and
close symbols for field-cooled data. (b) Field dependence of FM
magnetization at room temperature (the diamagnetic and
paramagnetic signal are subtracted). } \label{fig3}
\end{figure}

Temperature and field dependence of magnetization of 40 nm films
is shown in Fig. \ref{fig3}. The as-grown film A1 shows a FM
behavior below 350 K, which is also higher than that in bulk
material. In GBCO bulk material, the annealing at 400 $\celsius$
in 1 bar oxygen will increase the oxygen content (5+$\delta$) to a
higher value ($>$5.53), which suppresses the FM order to lower
temperature\cite{Taskin}. In film, this effect is even much
notable: the spontaneous magnetization in A2 disappears above 77K.
Either higher sensitivity of FM order to oxygen content in film,
or higher oxygen content in film, should be responsible to this.
The annealing at 900 $\celsius$ in air and cooling to room
temperature with furnace is believed to reduce the oxygen content
to $\sim$5.5 in bulk materials\cite{Taskin}, but in film, this
procedure has another effect: relaxation of strain comes from
substrate. In A3, the spontaneous magnetization recovers, and the
curie temperature shift to higher than 400 K. The x-ray
diffraction pattern in Fig. \ref{fig1} also shows an additional
peak comparing to A1 and A2, which is the result of strain
relaxation. The M(H) curves in Fig. \ref{fig3} (b) confirm the
remnant magnetization above room temperature in A1 and A3, and
also the disappearance of remnant magnetization above room
temperature in A2.

\begin{figure}[ht]
\centering
\includegraphics[width=8cm]{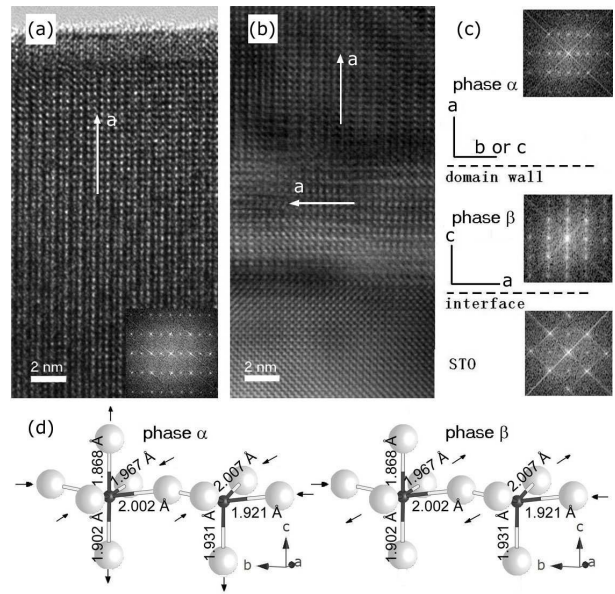}
\caption{(a) HRTEM image taken from the surface of 100 nm film.
inset shows the FFT result of this area. (b) HRTEM image taken
from the interface of A3. Two domains can be clearly seen in the
film. The FFT results of different areas in (b) are shown in (c).
(d) The strain state in phase $\alpha$ and $\beta$. } \label{fig4}
\end{figure}

Fig.\ref{fig4} (a) shows the HRTEM cross-section image taken from
the surface of 100 nm film. Doubled lattice period is observed in
in-plane direction but not in out-of-plane direction. The fast
Fourier transform (FFT) result of this area, shown in the inset of
Fig.\ref{fig4} (a), confirms this observation. It has been
mentioned above that in GBCO only $a$-axis is not doubled
comparing with perovskite lattice, which enable us to identify
different axis in GBCO film. So the $a$-axis of GBCO film in Fig.
\ref{fig4} (a) is out of plane. Some literatures also reported the
superstructure along $a$-axis due to atom
displacement,\cite{Chernenkov} but the intensity of superstructure
reflection they observed is so low (2$\sim$4 orders lower than
fundamental ones) that we won't see it even in our FFT results.
Fig. \ref{fig4} (b) shows the HRTEM cross-section image of A3,
where two domains are present in film. For the domain near the
substrate-film interface, the in-plane lattice is not doubled, but
for the domain near the film surface, the out-of-plane lattice is
not doubled. The FFT results of different areas are shown in Fig.
\ref{fig4} (c). For the same reason mentioned above, we can
identify $a$-axis of different domains, which is schematic shown
in Fig.\ref{fig4} (c). In fact, the domain with in-plane $a$-axis
is also found near the interface of 100nm film and in the whole
film of A1, but the domain with out-of-plane $a$-axis is not found
in A1. The HRTEM results are consistent with XRD patterns in Fig.
\ref{fig1}, if we regard phase $\alpha$ to the domain with
out-of-plane $a$-axis and phase $\beta$ to the domain with
in-plane $a$-axis.

One may ask that why phase $\alpha$ presents, since $a$-axis is
much closer to the lattice parameter of STO substrate than
$c$-axis. Here is a phenomenology explanation. If we just consider
the lattice parameter, $c$-axis will surely present out-of-plane
in order to reduce the elastic energy of film, since $a$ = 3.88
\AA \mbox{ }and $b$/2 = 3.92 \AA \mbox{ }is much closer to the STO
lattice parameter (3.906 \AA) comparing with $c$/2 = 3.77 \AA. But
as mentioned by Taskin $et$ $al$.\cite{Taskin}, the oxygen atom
diffusion along $c$-axis is negligible in comparison of $ab$ plane
in GBCO. And the equilibrium oxygen content is very sensitive to
the temperature and oxygen pressure. During the deposition and
annealing, different heat treatments are surely applied to the
film. $a$ or $b$-axis lies out-of-plane is a good choice to favor
the oxygen diffusion during different heat treatment. But $b$ will
not present out-of-plane considering that $b$-axis is the only one
larger than that of STO substrate. So, in A1, only $\beta$ phase
is present due to strain astriction comes from substrate; but in
100 nm film, phase $\beta$ is present near the interface and phase
$\alpha$ is present near the surface of film due to the relaxation
of strain with increasing thickness. Phase $\alpha$ is also
present in A3, where the high-temperature annealing may enhance
the relaxation of strain by creating more defects. The defects can
be seen near the domain boundary, and even in phase $\beta$.

Now we should explain the origin of enhanced FM order, which is
expanded to both low temperature and high temperature. Based on
the XRD results in Fig. \ref{fig1}, $a$ and $b$ are compressed ,
$c$ is elongated in phase $\alpha$; $a$ is elongated and $b$ is
compressed, but $c$ is almost unchanged in phase $\beta$. These
strain state are indicated by arrows in Fig. \ref{fig4} (d), where
only CoO$_{6}$ octahedron and CoO$_{5}$ pyramid are present. The
bond length are reproduced from the structure information in
literature.\cite{Frontera2} First, about the disappearance of AFM
phase at low temperature. In bulk material, the AFM-FM transition
is caused by the appearance of $e_{g}$ electron of Co$^{3+}$ in
octahedra, which is induced by either the appearance of
Co$^{2+}$/Co$^{4+}$ pairs or the LS to IS/HS transition of
Co$^{3+}$ in some of the octahedra with increasing thermal
excitation. But in film, as shown in Fig. \ref{fig4} (d), the
strain in phase $\alpha$ and phase $\beta$ will tune the octahedra
to less distorted one, decrease the crystal-field splitting energy
$\Delta$$E$ and stabilize higher spin state of Co$^{3+}$ in it.
The appearance of $e_{g}$ electron of Co$^{3+}$ in octahedra will
surely kill the AFM phase\cite{Taskin} in all the films. Second,
about the enhanced FM order at high temperature. In bulk material,
the AFM orbital order in two-leg ladders induces FM order inside
ladders due to Goodenough-Kanamori rules. But these ladders will
be killed by increasing thermal fluctuation, which is the reason
of FM order disappears at high temperature. In film, we can't
identify the local distortion of pyramids (including the bond
length and bond angle) in this study, but our experimental results
indicate that the biaxial strain on the pyramids, as shown in Fig.
\ref{fig4} (d), has positive effect on the orbital ordering in
these two-leg ladders. Especially the strain in phase $\alpha$ can
increase the orbital-ordering temperature to a much higher one.
Enhanced FM order to high temperature was also observed in GBCO
films on (LaAlO$_{3}$)$_{0.3}$(Sr$_{2}$AlTaO$_{6}$)$_{0.7}$ with
$a$ = 3.86 \AA, where the same strain type as phase $\alpha$ can
be found (data not shown here). Pressure effect on the magnetic
properties of GBCO single crystal is needed to confirm our
supposition.

\section{CONCLUSION}

In summary, epitaxial GBCO films were grown on STO (001) single
crystal substrates. XRD and TEM patterns reveal the phase
separation phenomena in film. The magnetic behavior of the films
shows large difference from the bulk material: AFM-FM transition
disappears and Curie temperature is shifted to a much higher
temperature. The origin of this enhanced FM order is due to the
substrate-induced strain, and a-oriented phase contributes much
more to the enhanced FM order.

\section*{ACKNOWLEDGMENTS} This work was supported by the Hong Kong
Polytechnic University Postdoctoral Scheme (G-YX0C) and internal
funding (1-BB84). The support from the Center for Smart Materials
is also acknowledged.\\


\begin{references}


\bibitem{Briceno}
G. Briceno, H. Y. Chang, X. D. Sun, P. G. Schultz, and X. D. Xiang
, Science {\bf 270}, 273(1995).

\bibitem{Terasaki}
I. Terasaki, Y. Sasago, and K. Uchinokura, Phys. Rev. B {\bf 56},
R12685(1997).

\bibitem{Senaris}
M. A. Senaris-Rodriguez and J. B. Goodenough, J. Solid State Chem.
{\bf 116}, 224(1995).

\bibitem{Tsubouchi}
S. Tsubouchi, T. Ky\^omen, M. Itoh, P. Ganguly, M. Oguni, Y.
Shimojo, Y. Morii, and Y. Ishii, Phys. Rev. B {\bf 66},
052418(2002).

\bibitem{Fujita}
T. Fujita, T. Miyashita, Y. Yasui, Y. Kobayashi, M. Sato, E.
Nishibori, M. Sakata, Y. Shimojo, N. Igawa, Y. Ishii, K. Kakurai,
T. Adachi, Y. Ohishi, and M. Takata, J. Phys. Soc. Jpn. {\bf 73},
1987(2004).

\bibitem{Fita}
I. Fita, R. Szymczak, R. Puzniak, I. O. Troyanchuk, J.
Fink-Finowicki, Ya. M. Mukovskii, V. N. Varyukhin, and H.
Szymczak, Phys. Rev. B {\bf 71}, 214404(2005).

\bibitem{Frontera}
C. Frontera, A. Caneiro, A. E. Carrillo, J. Or\'o-Sol\'e and J. L.
Garc\'ia-Mu\~noz, Chem. Mater. {\bf 17}, 5439(2005); C. Frontera,
J. L. Garc\'ia-Mu\~noz, A. E. Carrillo, M. A. G. Aranda, I.
Margiolaki, A. Caneiro, Phys. Rev. B {\bf 74}, 054406(2006).

\bibitem{Maignan}
A. Maignan, C. Martin, D. Pelloquin, N. Nguyen, and B. Raveau, J.
Solid State Chem. {\bf 142}, 247(1999).

\bibitem{Roy}
S. Roy, I. S. Dubenko, M. Khan, E. M. Condon, J. Craig, N. Ali, W.
Liu, B. S. Mitchell, Phys. Rev. B {\bf 71}, 024419(2005).

\bibitem{Pomjakushina}
E. Pomjakushina, K. Conder, and V. Pomjakushin, Phys. Rev. B {\bf
73}, 113105(2006).

\bibitem{Motin}
Md. Motin Seikh, Ch. Simon, V. Caignaert, V. Pralong, M. B.
Lepetit, S. Boudin, and B. Raveau, Chem. Mater. {\bf 20},
231(2008).

\bibitem{Khalyavin}
D. D. Khalyavin, D. N. Argyriou, U. Amann, A. A. Yaremchenko and
V. V. Kharton, Phys. Rev. B {\bf 75}, 134407(2007).

\bibitem{Troyanchuk}
I. O. Troyanchuk, N. V. Kasper, D. D. Khalyavin, H. Szymczak, R.
Szymczak, and M. Baran, Phys. Rev. Lett. {\bf 80}, 3380(1998).

\bibitem{Respaud}
M. Respaud, C. Frontera, J. L. Garc\'ia-Mu\~noz, Miguel \'Angel G.
Aranda, B. Raquet, J. M. Broto, H. Rakoto, M. Goiran, A. Llobet,
and J. Rodr\'iguez-Carvajal, Phys. Rev. B {\bf 64}, 214401(2001).

\bibitem{Taskin}
A. A. Taskin, A. N. Lavrov, and Y. Ando, Phys. Rev. Lett. {\bf
90}, 227201(2003); A. A. Taskin, A. N. Lavrov, and Y. Ando, Phys.
Rev. B {\bf 71}, 134414(2005); A. A. Taskin, A. N. Lavrov, and Y.
Ando, Phys. Rev. B {\bf 73}, 121101(R)(2006); A. A. Taskin, A. N.
Lavrov, and Y. Ando, Progress in Solid State Chem. {\bf 35},
481(2007).

\bibitem{Wu}
X. S. Wu, H. L. Zhang, J. R. Su, C. S. Chen, and W. Liu, Phys.
Rev. B {\bf 76}, 094106(2007).

\bibitem{Kasper}
N. V. Kasper, P. Wochner, A. Vigliante, H. Dosch, G. Jakob, H. D.
Carstanjen, and R. K. Kremer, J. Appl. Phys. {\bf 103},
013907(2008).

\bibitem{Yuan}
Z. Yuan, J. Liu, C. L. Chen, C. H. Wang, X. G. Luo, X. H. Chen, G.
T. Kim, D. X. Huang, S. S. Wang, A. J. Jacobson, and W. Donner,
Appl. Phys. Lett. {\bf 90}, 212111(2007).

\bibitem{Frontera2}
C. Frontera, J. L. Garc\'ia-Mu\~noz, A. Llobet, and M. A. G.
Aranda, Phys. Rev. B {\bf 65}, 180405(R)(2002).

\bibitem{Chernenkov}
Yu. P. Chernenkov, V. P. Plakhty, V. I. Fedorov, S. N. Barilo, S.
V. Shiryaev, and G. L. Bychkov, Phys. Rev. B {\bf 71},
184105(2005).









\newpage

\noindent

\end{references}
\end{document}